\documentclass[12pt,epsf]{article}
\usepackage{graphicx}
\usepackage{amssymb}

\begin{document}

\def\a{\alpha}
\def\b{\beta}
\def\c{\varepsilon}
\def\d{\delta}
\def\e{\epsilon}
\def\f{\phi}
\def\g{\gamma}
\def\h{\theta}
\def\k{\kappa}
\def\l{\lambda}
\def\m{\mu}
\def\n{\nu}
\def\p{\psi}
\def\q{\partial}
\def\r{\rho}
\def\s{\sigma}
\def\t{\tau}
\def\u{\upsilon}
\def\v{\varphi}
\def\w{\omega}
\def\x{\xi}
\def\y{\eta}
\def\z{\zeta}
\def\D{\Delta}
\def\G{\Gamma}
\def\H{\Theta}
\def\L{\Lambda}
\def\F{\Phi}
\def\P{\Psi}
\def\S{\Sigma}
\def\BR{{\rm Br}}
\def\o{\over}
\def\beq{\begin{eqnarray}}
\def\eeq{\end{eqnarray}}
\newcommand{\nn}{\nonumber \\}
\newcommand{\gsim}{ \mathop{}_{\textstyle \sim}^{\textstyle >} }
\newcommand{\lsim}{ \mathop{}_{\textstyle \sim}^{\textstyle <} }
\newcommand{\vev}[1]{ \left\langle {#1} \right\rangle }
\newcommand{\bra}[1]{ \langle {#1} | }
\newcommand{\ket}[1]{ | {#1} \rangle }
\newcommand{\EV}{ {\rm eV} }
\newcommand{\KEV}{ {\rm keV} }
\newcommand{\MEV}{ {\rm MeV} }
\newcommand{\GEV}{ {\rm GeV} }
\newcommand{\TEV}{ {\rm TeV} }
\def\diag{\mathop{\rm diag}\nolimits}
\def\Spin{\mathop{\rm Spin}}
\def\SO{\mathop{\rm SO}}
\def\O{\mathop{\rm O}}
\def\SU{\mathop{\rm SU}}
\def\U{\mathop{\rm U}}
\def\Sp{\mathop{\rm Sp}}
\def\SL{\mathop{\rm SL}}
\def\tr{\mathop{\rm tr}}

\newcommand{\bear}{\begin{array}}  
\newcommand {\eear}{\end{array}}
\newcommand{\la}{\left\langle}  
\newcommand{\ra}{\right\rangle}
\newcommand{\non}{\nonumber}  
\newcommand{\ds}{\displaystyle}
\newcommand{\red}{\textcolor{red}}
\def\ubl{U(1)$_{\rm B-L}$}
\def\REF#1{(\ref{#1})}
\def\lrf#1#2{ \left(\frac{#1}{#2}\right)}
\def\lrfp#1#2#3{ \left(\frac{#1}{#2} \right)^{#3}}
\def\OG#1{ {\cal O}(#1){\rm\,GeV}}

\def\TODO#1{ {\bf ($\clubsuit$ #1 $\clubsuit$)} }


\baselineskip 0.7cm

\begin{titlepage}

\begin{flushright}
IPMU-12-0133\\
YITP-12-52
\end{flushright}

\vskip 1.35cm
\begin{center} 
{\large \bf Higgs Boson Mass in Low Scale Gauge Mediation Models
}
\vskip 1.2cm

{Tsutomu T. Yanagida$^1$, Norimi Yokozaki$^1$ and Kazuya Yonekura$^2$}

\vskip 0.4cm

{\it $^1$ Kavli Institute for the Physics and Mathematics of 
the Universe (IPMU),\\ 
University of Tokyo, Chiba 277-8583, Japan\\
$^2$ Yukawa Institute for Theoretical Physics, Kyoto University,\\ Kyoto 606-8502, Japan
}

\vskip 1.5cm

\abstract{
We consider low scale gauge mediation models with a very light gravitino $m_{3/2} \lesssim 16$~eV, in the light of recent experimental hints on the Higgs boson mass. The light gravitino is very interesting since there is no gravitino over-production problem, but it seems difficult to explain
the Higgs boson mass of $\sim125$ GeV. This is because of the conflict between the light gravitino mass and heavy SUSY particle masses needed for producing the relatively heavy Higgs boson mass.
We consider two possible extensions in this paper: 
a singlet extension of the Higgs sector, and strongly coupled gauge mediation. 
We show that there is a large parameter space, in both scenarios, where the Higgs boson mass of $\sim$125 GeV is explained without any conflict with such a very light gravitino.
}
\end{center}
\end{titlepage}

\setcounter{page}{2}
\section{Introduction}
Gauge mediated SUSY breaking models~\cite{Dine:1981za,mGMSB} are attractive since there is no flavor changing neutral current problem, and the SUSY CP problem can be easily solved~\cite{SUSYCP0, SUSYCP, Hamaguchi1, Hamaguchi2}.~\footnote{A non-negligible CP violating phase may arise from supergravity effects and GUT breaking effects~\cite{Moroi:2011fi}.}  Among gauge mediation models, low scale gauge mediation models are interesting from the cosmological point of view; if the gravitino is as light as $m_{3/2} \lesssim 16$~eV, there is no gravitino over-production problem~\cite{Viel:2005qj}. In such low scale gauge mediation models with the very light  gravitino, the observed baryon asymmetry is successfully explained by the thermal leptongenesis~\cite{Fukugita:1986hr}, since there is no upper bound for the reheating temperature. 

On the other hand, recent LHC results suggest that the mass of the standard-model (SM) like Higgs boson is near $125~\GEV$ for both CMS~\cite{Chatrchyan:2012tx} and ATLAS~\cite{ATLAS_Higgs} experiments. Such a relatively heavy Higgs boson requires sparticle mass of $\mathcal{O}(10)$ TeV~\cite{Okada:1990gg} if there is no large trilinear coupling of the stops, $A_t$, as in the case of 
conventional gauge mediation models.
In weakly coupled gauge mediation models, $\mathcal{O}(10)$ TeV sparticles are not realized consistently with $m_{3/2} \lesssim 16$ eV, where 
$m_{3/2} \gtrsim ({F}/{\sqrt{3} M_P})$, since the upper bound on the SUSY breaking parameter $F$ leads to an upper bound on the sparticle masses of 
${\cal O}(1)~\TEV$.

A number of extended gauge mediation models have been proposed 
which can explain the Higgs boson mass of $125$~GeV with $\mathcal{O}(1)$ TeV sparticles.
Recent studies include: gauge mediation models with vector-like matters~\cite{Endo:2011mc, Evans:2011uq, Endo:2011xq, Nakayama:2012zc, Martin:2012dg}, the $Z_3$ symmetric singlet extension of the Higgs sector~\cite{Hamaguchi1, Larsen:2012rq} (earlier works can be found in Refs.~\cite{Ellwanger:2008py, Morrissey:2008gm}), introducing $U(1)'$ gauge symmetry (the Higgs is charged under this new $U(1)'$ symmetry)~\cite{Endo:2011gy}, generation of large $A_t$ through Higgs-Messenger mixing~\cite{Evans:2011bea, Evans:2012hg,Kang:2012ra,Craig:2012xp} or through RG flow~\cite{Draper:2011aa}, 
introducing extra strongly interacting sector which couples to the Higgs sector~\cite{Heckman:2011bb,Evans:2012uf,Kitano:2012wv}, and so on.
However, it is not easy to accommodate $m_{3/2} \lesssim 16$ eV in many of these models. 
For example, if we introduce additional particles charged under the SM gauge groups,
the perturbative unification allows us to introduce only a small number of messengers.
However, such a small messenger number may be excluded by the constraints from recent LHC SUSY search~\cite{CMS_SUSY}
and a vacuum stability condition with very light gravitino~\cite{Hisano:2007gb,Hisano:2008sy}. 

In this paper, we discuss the Higgs boson mass in two possible extensions of gauge mediation, which can be consistent with $m_{3/2} \lesssim 16$ eV
(and also with the hidden sector dark matter): 
the singlet extension of the Higgs sector, and strongly coupled gauge mediation.
In the former case, 
there is an additional tree level contribution to the Higgs boson mass from superpotential. Therefore, $m_h \simeq 125$ GeV can be explained without any conflict with the ultra light gravitino, in principle. As we will explain in the next section, however, it turns out to be difficult to explain the Higgs boson mass in the $Z_3$ invariant model, i.e, Next-to-minimal supersymmetry standard model (NMSSM)~\cite{NMSSM_review}, while keeping the perturbative unification. We show that it is  possible to explain $m_h \simeq 125$ GeV in a more general singlet extension of the Higgs sector, where all couplings are kept in a perturbative regime.
In this case, we have $\mathcal{O}(1)$ TeV sparticles.

Another possibility we discuss in this paper is the case of strongly coupled gauge mediation. In this case, the sfermion masses of $\mathcal{O}(10)$ TeV may be realized even in the region with $m_{3/2} \lesssim 16$ eV as we will discuss in Sec.~\ref{sec:stronglycoupled}.  
We also point out that the quartic coupling of the Higgs potential which is usually determined by the gauge couplings, can be deviated due to contributions from messenger loops. This effect possibly increases the Higgs boson mass by $\mathcal{O}(1)$ GeV in low scale gauge mediation. 
Therefore we argue that the soft SUSY breaking mass scale can be a few TeV to obtain the 125 GeV Higgs boson mass with the help of this effect.

This paper is organized as follows. In section~\ref{sec:singlet}, we discuss singlet extended models in gauge mediation. 
We show that a general model of the singlet extension can explain the Higgs boson mass of 125 GeV. In section~\ref{sec:stronglycoupled}, gauge mediation with strongly coupled messengers are discussed. We pointed out two possibilities which increase the Higgs boson mass:
large sfermion masses of order $\mathcal{O}(10)$ TeV, and the deviation of the D-term coefficient from the gauge couplings. Section~\ref{sec:conc} is devoted to summary and discussion. 

\section{Singlet Extension  of the Gauge Mediation Models}\label{sec:singlet}

In MSSM, the mass of the lightest Higgs boson is constrained as $m_h^0 \lesssim m_Z$ at the tree level, since the quartic coupling of the Higgs is completely
determined by the gauge couplings of $SU(2)_L$ and $U(1)_Y$. However, additional contribution can arise through superpotential if a gauge singlet superfield, $S$, is introduced to the Higgs sector as
\begin{eqnarray}
W = \lambda S H_u H_d,
\end{eqnarray} 
where $H_u$ and $H_d$ are up-type Higgs and down-type Higgs, respectively. Then,
the tree level upper bound for the Higgs boson mass is relaxed as~\cite{Drees:1988fc}
\begin{eqnarray}
m_{h^0}^2 \leq m_Z^2 \left( \cos^2 2\beta  + \frac{2 \lambda^2}{g_Y^2 + g^2} \sin^2 2\beta \right), \label{eq:tree}
\end{eqnarray}
where $\beta$ is defined by the ratio of the vacuum expectation values of $H_u$ and $H_d$ as $\tan\beta=\left<H_u^0\right>/\left<H_d^0\right>$. We denote the gauge couplings of $U(1)_Y$ and $SU(2)$ as $g_Y$ and $g$, respectively. The second term comes from the F-term contribution of the singlet $S$
to the Higgs potential, which is proportional to $\lambda^2$ and suppressed when $\tan\beta$ is large. It is possible to raise the tree level Higgs boson mass above $m_Z$ when $\lambda^2$ is larger than $(g^2+g_Y^2)/2$; 
 the Higgs boson mass $m_{h} \simeq 125$~GeV can be explained even when the stop mass is $\mathcal{O}(1) $ TeV and the trilinear coupling of the stop, $A_t$ is not large. In fact, $A_t$ is zero at the messenger scale in minimal gauge mediation models and it is generated through renormalization group running between the messenger scale and the soft mass scale.

Among singlet extensions, a well-known model is the so-called Next-to-Minimal Supersymmetric Standard Model (NMSSM), which has $Z_3$ invariant potential~\cite{NMSSM_review}. The superpotential and the soft SUSY breaking terms are assumed as
\begin{eqnarray}
W &=& \lambda S H_u H_d + \frac{\kappa}{3} S^3, \nonumber \\
 V_{\rm soft} &=& m_S^2 |S|^2 + m_{H_u}^2 |H_u|^2 + m_{H_d}^2 |H_d|^2 + (A_\lambda \lambda S H_u H_d + A_\kappa \frac{\kappa}{3} S^3 + h.c.).
\end{eqnarray}
In the NMSSM, the Higgsino mass parameter, $\mu$, is given by the VEV of the singlet field $S$ as $\lambda \left< S \right>$; the electroweak scale $\mu$ parameter requires large enough VEV of $S$, which is induced by negative $m_S^2$ of $\mathcal{O}(1~{\rm TeV}^2)$. It is known that such a negative $m_S^2$ is difficult to be obtained within (minimal) gauge mediation models. When the extra vector-like matters which couple to $S$ are introduced, such a negative $m_S^2$ of $\mathcal{O}(1~{\rm TeV}^2)$ can be obtained through renormalization group evolution~\cite{Dine:1994vc, deGouvea:1997cx}.
Note that $Z_3$ symmetry can be anomalous due to the existence of the extra matters and 
the cosmological domain wall problem associated with spontaneous break down of the discrete symmetry can be solved~\cite{Preskill:1991kd}. In particular, if the extra vector-like matters are charged under a hidden strong gauge group, the domain wall problem is solved without disturbing the Peccei Quinn solution to the strong CP problem~\cite{Hamaguchi1, Hamaguchi2}. 

However even though the weak scale $\mu$ parameter is obtained, it is still difficult to explain the Higgs boson  mass, $m_{h^0} \simeq 125$ GeV; 
the Higgs boson mass $m_{h^0}$ crucially depends on the mixings among $H_u$, $H_d$ and $S$. If the mixings are not small, even smaller value of $m_{h^0}$ than that of MSSM is predicted.
In the singlet extended models, the mass matrix of the CP even Higgs is extended to $3\times 3$ as 
\begin{eqnarray}
-\mathcal{L}_{\rm mass}=\frac{1}{2} \left( h_d\  h_u\  h_s \right)
\left(
\begin{array}{ccc}
M_{dd}^2 & M_{du}^2 & M_{ds}^2 \\
  & M_{uu}^2 & M_{us}^2 \\
  &  & M_{ss}^2 
\end{array}
\right)
\left(
\begin{array}{c}
h_d \\
h_u \\
h_s
\end{array}
\right),
\end{eqnarray}
where $h_d$, $h_u$ and $h_s$ are the fluctuations around the VEV of  $H_d$, $H_u$ and $S$, respectively. 
The mass eigenstates are related to $h_u$, $h_d$ and $h_s$ as
\begin{eqnarray}
\left(
\begin{array}{c}
h_1 \\
h_2 \\
h_3
\end{array}
\right)=
\left(
\begin{array}{ccc}
O_{1d} & O_{1u} & O_{1s} \\
O_{2d}  & O_{2u} & O_{2s} \\
O_{3d}  & O_{3u} & O_{3s} 
\end{array}
\right)
\left(
\begin{array}{c}
h_d \\
h_u \\
h_s
\end{array}
\right),
\end{eqnarray}
where $h_1$ is the SM-like Higgs. In order to raise the Higgs boson mass, $h_1$, a mixing parameter, $O_{1s}$, should be small. Such a small mixing is realized when  $M_{us}^2$ and $M_{ds}^2$ are small enough compared to $M_{uu}^2, M_{dd}^2$ and $M_{du}^2$, or when $M_{ss}^2$ is large compared to the upper 2x2 matrix elements. 
However, in the NMSSM, it is difficult to achieve such a small mixing within gauge mediation models, in general.
The matrix elements, $M_{us}^2$, $M_{ds}^2$ and $M_{ss}^2$ are given by 
\begin{eqnarray}
M_{us}^2 &\simeq&\lambda v_u (2\mu_{\rm eff}-(A_\lambda + 2 \kappa v_S )/\tan\beta), \nonumber \\
M_{ds}^2 &\simeq&\lambda v_d (2\mu_{\rm eff}-(A_\lambda + 2 \kappa v_S )\tan\beta), \nonumber \\
M_{ss}^2 &\simeq& \lambda A_\lambda v^2 \sin\beta \cos\beta/v_S + \kappa v_S (A_\kappa + 4\kappa v_S),
\end{eqnarray}
where $\mu_{\rm eff}$ is $\lambda v_S$ and the radiative corrections are neglected.
Here $m_{H_u}^2$, $m_{H_d}^2$ and $m_S^2$ are written in terms of other parameters by using three minimization conditions.
Since $A_\lambda$ is not large in gauge mediation models, the small mixings are realized only when $\kappa \gtrsim \lambda$. 

\begin{figure}[t]
\begin{center}
\includegraphics[width=7.9cm]{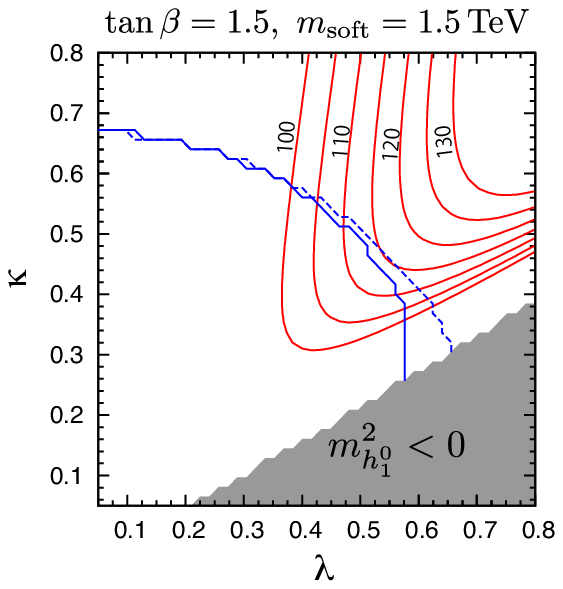}
\includegraphics[width=7.9cm]{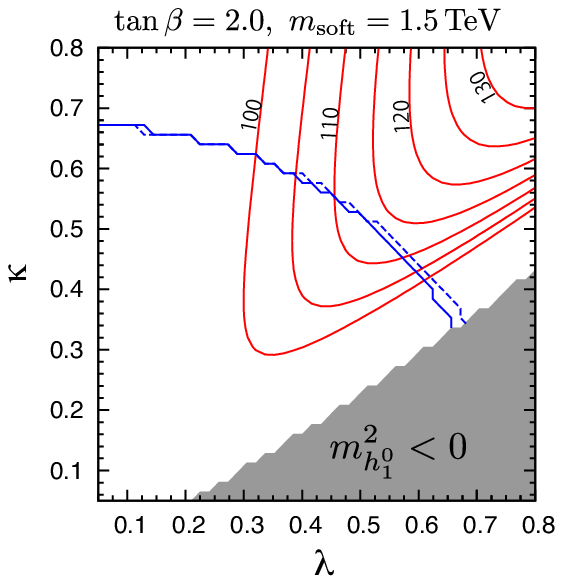}
\caption{The contours of the Higgs boson mass in $\lambda$-$\kappa$ plane. Here we take $A_\kappa=-50$ GeV and other scalar trilinear couplings are taken as zero. The blue solid (dashed) line shows the upper bound of $\lambda$ and $\kappa$ for $N_{5}=0 ~ (N_5=4)$ at the mass scale, $M=10^5$ GeV. We take $\mu_{\rm eff}=500$ GeV.}
\label{fig:l_k_higgs}
\end{center}
\end{figure}

In Fig.~{\ref{fig:l_k_higgs}}, the contours of the Higgs boson mass in $\lambda-\kappa$ plane are shown. The calculations are performed by using {\tt NMHDECAY}~\cite{NMHDECAY}. The soft masses (gaugino and squark masses) are taken as $m_{\rm soft}=1.5$ TeV and the trilinear coupling of the stop, $A_t$, is set to $A_t=0$. The other trilinear couplings, $A_\lambda$ and $A_\kappa$, are set to $A_\lambda=0$ and $A_\kappa=-50$ GeV, respectively.
The region outside the blue solid line are excluded from the Landau pole constraint on $\lambda$ and $\kappa$; the perturbativity up to the GUT scale is imposed on $\lambda$ and $\kappa$.~\footnote{However, it is possible to UV-complete much larger values of $\lambda$~\cite{Nakayama:2011iv}.} 
In the case that the extra-matters (e.g. messengers) exist, the constraint is relaxed. This can be shown from the beta function of $\lambda$, which is given by
\begin{eqnarray}
16\pi^2 \frac{d \lambda}{d \ln Q} \simeq \lambda(3Y_t^2+4\lambda^2+2\kappa^2-g_Y^2-3g_2^2)
\end{eqnarray}
where $Q$ is the renormalization scale and $Y_t$ is the top Yukawa coupling. Since the gauge couplings contribute to $d \lambda/(d\ln Q)$ negatively, the upper bound on $\lambda$ is somewhat relaxed when the gauge couplings are large. Such large gauge couplings are realized when the extra matters which have SM charge exist. The blue dashed line shows the upper bound on $\lambda$ and $\kappa$ in the case that 4 pairs of the extra-matters (${\bf 5}$ and ${\bf \bar{5}}$ under $SU(5)$ gauge group) exist at the mass scale, $M=10^5\, {\rm GeV}$. Unfortunately, even if 4 pairs of the extra-matters exist, the Higgs boson mass of 125 GeV  can not be explained in the region consistent with the Landau-pole constraint.~\footnote{
The constraint from the vacuum stability may also be important, depending on the parameters in the Higgs sector~\cite{Kanehata:2011ei}.}

Alternatively, we consider a more generic superpotential~\footnote{By a shift of the $S$ field, 
this superpotential becomes very similar to the one studied in Ref.~\cite{Dine:2007xi}.}
\begin{eqnarray}
W = \lambda S H_u H_d  + \mu' S^2/2 + \xi_F S + \kappa S^3/3. \label{eq:S_super}
\end{eqnarray}
Here, we assume the coefficient of the cubic term, $\kappa$, is small and we neglect $\kappa S^3/3$ in the following analysis. The soft SUSY breaking terms are given by,
\begin{eqnarray}
-\mathcal{L}_{\rm soft}&=& m_S^2 |S|^2 + m_{H_u}^2 |H_u|^2 + m_{H_d}^2 |H_d|^2 \nonumber \\
&+& (A_\lambda \lambda S H_u H_d  + B' \mu' S^2/2 + \xi_S S) + h.c..
\end{eqnarray}
The dimensionful parameters in the superpotential, $\mu'$ and $|\xi_F|^{1/2}$ are of the order of TeV.
The soft mass parameters, $m_{H_u}^2$ and $m_{H_d}^2$, are determined by the (minimal) gauge mediated SUSY breaking model, while $m_S^2$, $B'$, vanish at the messenger scale. The linear term, $\xi_S$ of $\mathcal{O}({\rm TeV
})$ is generated radiatively, if $S$ couples to the messengers~\cite{Ellwanger:2008py}. 
With $\xi_S$ of the order of TeV$^3$, the correct electroweak symmetry breaking occurs without negative $m_S^2$ of TeV$^2$; a large enough vacuum expectation value of $S$ is obtained in gauge mediation models.

With such a superpotential and a scalar potential, $M_{us}^2$ and $M_{ds}^2$ are written by
\begin{eqnarray}
M_{us}^2 &\simeq&\lambda v_u (2\mu_{\rm eff}-(A_\lambda +  \mu' )/\tan\beta), \nonumber \\
M_{ds}^2 &\simeq&\lambda v_d (2\mu_{\rm eff}-(A_\lambda +  \mu' )\tan\beta), \nonumber \\ 
M_{ss}^2 &\simeq& m_S^2 + B' \mu'/2 + \mu'^2.  
\end{eqnarray}
Since $m_S^2$ and $B'$ are small in gauge mediation, $M_{ss}^2$ is dominated by $\mu'^2$.
When $\mu' \gtrsim 2\mu$, the mixing, $O_{1s}$ is effectively suppressed and the Higgs boson mass is raised by 
the F-term contribution to the Higgs potential.

Note that TeV scale $|\xi_F|^{1/2}$ is also required for successful electroweak symmetry breaking. The minimization conditions lead to
\begin{eqnarray}
\frac{1}{2} \sin 2\beta = \frac{ B_{\rm eff} \mu_{\rm eff}}{m_{H_u}^2 + m_{H_d}^2 + 2 |\mu_{\rm eff}|^2 + \lambda^2 (v_u^2+v_d^2)}, \label{eq:ewsb1}
\end{eqnarray}
where $B_{\rm eff}=A_\lambda + \mu' + \xi_F/v_S$. Since $\mu' \gtrsim 2 \mu$, cancellation between $\xi_F/v_S$ and $\mu'$ is required to satisfy the above relation (\ref{eq:ewsb1}).

\begin{figure}[t]
\begin{center}
\includegraphics[width=7.9cm]{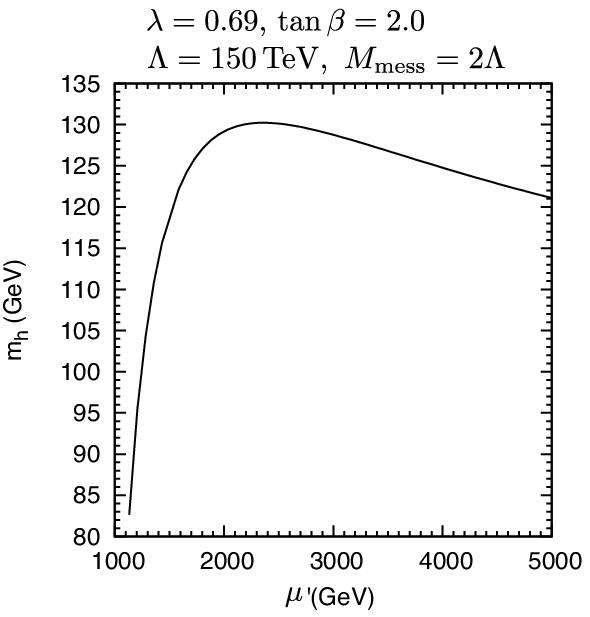}
\includegraphics[width=7.9cm]{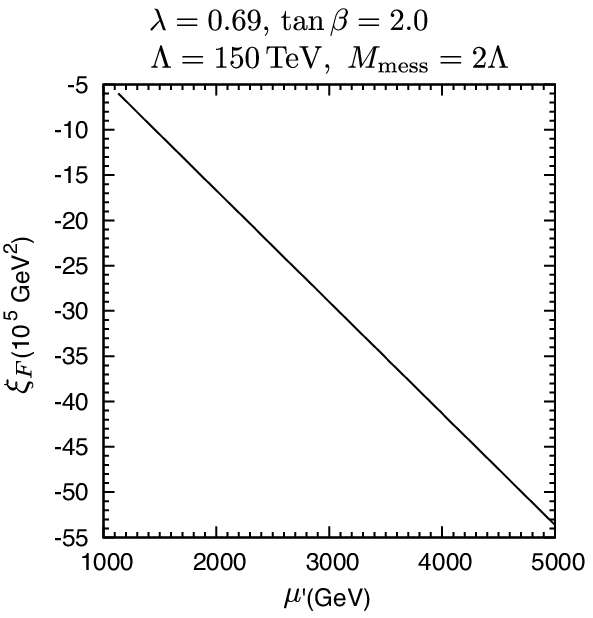}
\caption{The Higgs boson mass and $\xi_F(M_{\rm mess})$ as functions of $\mu'$. The messenger number is taken as $N_{\rm eff}=1$. Here we assume that the masses of sfermions and gauginos are those in the minimal gauge mediation model.}
\label{fig:singlet_res_n5_1}
\end{center}
\end{figure}

\begin{figure}[t]
\begin{center}
\includegraphics[width=7.9cm]{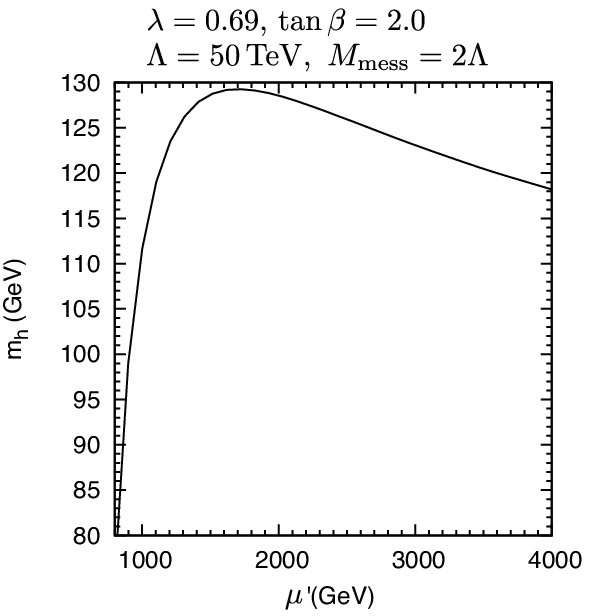}
\includegraphics[width=7.9cm]{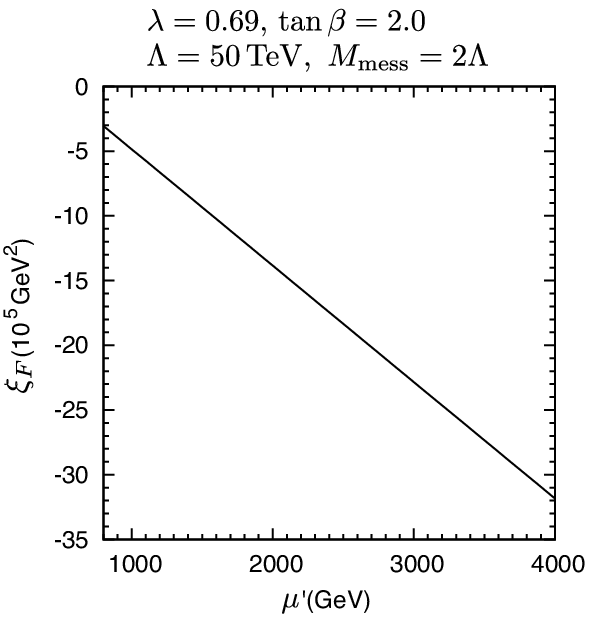}
\caption{The Higgs boson mass and $\xi_F(M_{\rm mess})$ as functions of $\mu'$. The messenger number is taken as $N_{\rm eff}=4$. Here we assume that the masses of sfermions and gauginos are those in the minimal gauge mediation model.}
\label{fig:singlet_res_n5_4}
\end{center}
\end{figure}

Results of numerical calculations are shown in Figures~\ref{fig:singlet_res_n5_1} and \ref{fig:singlet_res_n5_4} . The calculations are performed by using the {\tt NMGMSB}~\cite{Ellwanger:2008py}. 
In Fig.~\ref{fig:singlet_res_n5_1} and \ref{fig:singlet_res_n5_4}, the Higgs boson mass $m_h$ and $\xi_F$ are shown as functions of $\mu'$. (The linear term coefficient, $\xi_S$, is also determined by the minimization conditions.)
We have taken the parameters in the Higgs sectors as $\lambda=0.69$ and $\tan\beta=2.0$. 
The messenger number $N_{\rm eff}$ and the SUSY breaking parameter $\Lambda_{\rm eff}$ are taken to be $N_{\rm eff}=1$ and $\Lambda =150$ TeV for Fig.~\ref{fig:singlet_res_n5_1} and $N_{\rm eff}=4$ and $\Lambda =50$ TeV for Fig.~\ref{fig:singlet_res_n5_4}, respectively. The messenger scale is set to be $M_{\rm mess}=2\Lambda$. The squark mass and the gluino mass are $m_{\tilde{q}} \simeq 1.6 \, {\rm TeV}$ ($1.3$ TeV) and $m_{\tilde{g}} \simeq 1.2$ TeV ($1.5$ TeV) in Fig~\ref{fig:singlet_res_n5_1} (Fig.~\ref{fig:singlet_res_n5_4}). 
The Higgs boson mass of $125$ GeV can be explained with $\mu' \simeq 1.7\, {\rm TeV},\,  3.8\, {\rm TeV}\ (1.2\, {\rm TeV},\,  2.7\, {\rm TeV})$ in Fig.~\ref{fig:singlet_res_n5_1} (Fig.~\ref{fig:singlet_res_n5_4}). Corresponding $\mu$ parameters are $\mu_{\rm eff}\simeq 880$ GeV and $640$ GeV in Fig.~\ref{fig:singlet_res_n5_1} and Fig.~\ref{fig:singlet_res_n5_4}, respectively.  In the case that $N_{\rm eff}=1$, there is a sever constraint on the squark and the gluino masses from the vacuum stability~\cite{Hisano:2008sy}, while the constraint can be weaken for $N_{\rm eff}=4$. Note that the perturbativity of the Yukawa couplings up to the GUT scale is easily maintained.

\section{Higgs boson mass in strongly coupled gauge mediation models}\label{sec:stronglycoupled}
In minimal gauge mediation models, the gaugino masses and the sfermion masses are predicted to be of the same order 
as $m_{\rm gaugino} \sim m_{\rm sfermion}$, and trilinear couplings are generated only at 2-loop level.  
Therefore the Higgs boson mass, $m_{h} \simeq 125$ GeV, is explained only when all of the SUSY particles are as heavy as $\mathcal{O}(10)$ TeV. 
In weakly coupled models, such heavy SUSY particles may not be realized when the gravitino is as light as $m_{3/2} \lesssim 16~\EV$
(see Ref.~\cite{Ajaib:2012vc} for a recent discussion).

However, when the messenger fields are strongly coupled, the situation changes. 
In the following, we discuss advantages of strongly coupled low scale gauge mediation models.
\subsection{Heavy sfermions}
Let us recall a bound on the gravitino mass in minimal gauge mediation. 
Suppose that there are $N_{\rm mess}$ pairs of messenger fields $\Psi$ and $\tilde{\Psi}$
which are in the fundamental and anti-fundamental representations of $SU(5)_{\rm GUT} \supset SU(3)_C \times SU(2)_L \times U(1)$, respectively.
The superpotential is given as
\beq
W=(M_{\rm mess}+y Z)\tilde{\Psi}\Psi,
\eeq
where $Z$ is the SUSY breaking field with the vev given as $\vev{Z}=\theta^2 F$. All the parameters are taken to be real and positive without loss of generality.
Then, sparticle masses are roughly given as
\begin{eqnarray}
m_{\rm sfermion}^2 &\sim&  2 c_2(r)N_{\rm mess}\left(\frac{g^2}{16\pi^2}\right)^2  \Lambda^2, \nonumber \\
m_{\rm gaugino} &\sim& N_{\rm mess}\left(\frac{g^2}{16\pi^2}\right)   \Lambda,  \label{eq:minimalsparticlemass}
\end{eqnarray}
where $g$ represents the SM gauge couplings, $c_2(r)$ is the quadratic Casimir invariant of representation $r$ in which the sfermion transforms, and
$\Lambda = yF/M_{\rm mess}$. For the messenger fields not to be tachyonic, the SUSY breaking scale $F$ must satisfy the bound $yF < M_{\rm mess}^2$.
Then the gravitino mass is bound from below as
\beq
m_{3/2} =\frac{F}{\sqrt{3} M_{Pl}} >  
240~\EV \cdot \left(\frac{1}{y} \right)
 \cdot \left( \frac{\Lambda}{10^3~\TEV} \right)^2, \label{eq:boundgravitino}
\eeq
where $M_{Pl} \simeq 2.4 \times 10^{18}~\GEV$ is the reduced Planck scale. To realize the Higgs boson mass $m_h \simeq 125~\GEV$ by using only the top-stop
loop contribution, we need $\Lambda \sim {\cal O}(10^3)~\TEV$ as is shown later.
Thus the gravitino mass cannot satisfy the cosmological bound
$m_{3/2} \lesssim 16~\EV$ if the messenger is weakly coupled, i.e., $y \lsim {\cal O}(1)$.
However, as the coupling $y$ is increased, the bound on the gravitino mass becomes smaller.
If the messenger fields are strongly coupled so that $y$ is as large as $4\pi$, 
the bound (\ref{eq:boundgravitino}) may be compatible with $m_{3/2} \lesssim 16~\EV$.

Now we give an estimation for the sfermion and gravitino masses in more generic strongly coupled models.
Consider a strongly coupled model which has a single mass scale $\Lambda$. All the particles in the model are assumed to have 
masses of order $\Lambda$.
The SUSY is assumed to be broken by dynamics of some gauge group $G$ with color factor $N$ (e.g., $G=SU(N),~Sp(N)$, etc.).
For concreteness, let us further assume that there are fields $\Psi,~\tilde{\Psi}$ which are in the bi-fundamental and anti-bi-fundamental representation
of $G \times SU(5)_{\rm GUT}$, respectively. These fields play the role of messenger fields in gauge mediation. 
For explicit models realizing this situation, we have the conformal gauge mediation models
discussed in Refs.~\cite{Ibe:2007wp,Ibe:2008si,Izawa:2009nz,Yanagida:2010wf,Yanagida:2010zz} in mind.

In general gauge mediation (GGM)~\cite{Meade:2008wd}, sfermion masses are given as
\beq
m^2_{\rm sfermion}=c_2(r)\frac{g^4}{16\pi^2} \int_0^\infty d p^2 [ -3\tilde{C}_1(p^2)+4\tilde{C}_{1/2}(p^2)-\tilde{C}_0 (p^2) ], \label{eq:GGM}
\eeq
where $\tilde{C}_{0}(p^2),~\tilde{C}_{1/2}(p^2)$ and $\tilde{C}_1(p^2)$ are defined by the current correlators of the hidden sector 
(see Ref.~\cite{Meade:2008wd} for details).
In the high energy limit $p^2 \gg \Lambda$, these three functions take the same value, i.e. 
$
\tilde{C}_{0}(p^2)-\tilde{C}_{1/2}(p^2),~\tilde{C}_{0}(p^2)-\tilde{C}_{1}(p^2) \to 0~(p^2 \to \infty)
$
and the integrand of Eq.~(\ref{eq:GGM}) vanishes in this region. 
In the low energy limit $p^2 \ll \Lambda^2$,
they become independent of $p^2$, i.e., $\tilde{C}_{0}(p^2) \simeq \tilde{C}_{0}(0)$. Then we may have
\beq
m^2_{\rm sfermion} \sim c_2(r)\frac{g^4}{16\pi^2} \Lambda^2 [ -3\tilde{C}_1(0)+4\tilde{C}_{1/2}(0)-\tilde{C}_0 (0) ].
\eeq

If one applies naive dimensional analysis (NDA)~\cite{Manohar:1983,Georgi:1986,Georgi:1992dw,Luty:1997fk,Cohen:1997rt,Nishio:2012sk}, 
the current correlators may be estimated as (see e.g.~Ref.~\cite{Nishio:2012sk})
\beq
 -3\tilde{C}_1(0)+4\tilde{C}_{1/2}(0)-\tilde{C}_0 (0) \sim \frac{N}{8\pi^2}. \label{eq:sfermioncoll}
\eeq
Then, the sfermion masses are estimated as
\beq
m^2_{\rm sfermion} = 2Nc_2(r)\left(\frac{g^2}{16\pi^2} \right)^2 \Lambda^2. \label{eq:sfermionestimate}
\eeq
We define the precise value of $\Lambda$ by this equation.\footnote{The masses of the hidden sector particles may be different 
from $\Lambda$ by $\mathcal{O}(1)$ factor.}

On the other hand, in NDA, the estimation of the SUSY breaking scale $F$ is given as
\beq
F =\frac{BN}{4\pi}\Lambda^2.
\eeq
where the parameter $B$ is a numerical factor of order 1. In NDA, $B$
may be just set to unity. However the precise value of $B$ has great phenomenological importance.
We obtain
\beq
m_{3/2} \simeq 20~\EV \cdot BN \cdot \left( \frac{\Lambda}{10^3~\TEV} \right)^2 \simeq 20~\EV \cdot B \cdot \left( \frac{m_{\tilde t}}{10~\TEV} \right)^2,
\eeq
where $m_{\tilde t}$ is the stop mass at the mediation scale.
Therefore, this type of models may or may not be consistent with the cosmological bound depending on the precise value of $B$.

A comment on the gaugino masses is desirable. We will discuss that the Higgs boson mass $m_h \simeq125~\GEV$ is achieved for the stop mass
of order $10~\TEV$. In the minimal gauge mediation, the gauginos (especially the gluino) also have masses of this order.
However, in non-minimal gauge mediation models, the gaugino masses tend to be somewhat smaller than the sfermion masses
even if the SUSY breaking scale and the messenger scale are comparable.\footnote{This suppression is due to 
numerical factors in the formulae of the gaugino masses. For evaluations of these factors in weakly coupled models,
see Ref.~\cite{Izawa:1997gs} for the case of a $F$-term direct gauge mediation model with
a stable SUSY breaking vacuum,
and Ref.~\cite{Shirai:2010rr} for a semi-direct gauge mediation model. }
Therefore, we have a possibility to discover the gauginos at the LHC even if the sfermions are too heavy.

\subsubsection*{Higgs boson mass}
Now let us discuss the Higgs boson mass. We assume that the sfermion masses are given by Eq.~(\ref{eq:sfermionestimate}) with $N(=N_{\rm mess})=4$
at the messenger scale, and the gaugino masses are $1~\TEV$.
Since the stop mass scale, defined as $M_S=\sqrt{m_{\tilde t_1}m_{\tilde t_2}}$, is much larger than $m_Z$, it is more appropriate to calculate the Higgs boson mass by solving the renormalization group equation for the Higgs quartic coupling~\cite{Okada:1990gg}, rather than evaluating Feynman diagrams, or equivalently, the effective potential. 
At the scale below $M_S$, the effective Lagrangian of the Higgs sector is given by
\begin{eqnarray}
V(H) = m^2 |H|^2 + \frac{\lambda}{2} |H|^4,
\end{eqnarray}
where $H$ is the light linear combination of the Higgs doublets, $H_u$ and $H_d$. 
The quartic coupling $\lambda$ should be matched at $M_S$ as
\begin{eqnarray}
\lambda(M_S) = \frac{1}{4}(g_Y^2(M_S) + g^2(M_S)) \cos^2 2\beta + \Delta \lambda, \label{eq:quartic}
\end{eqnarray}
where $g_Y$ and $g$ are the gauge couplings for $U(1)_Y$ and $SU(2)_L$ respectively.
At scales below $M_S$, the couplings $g_Y$, $g$ and $\lambda$ (and also the top Yukawa coupling $Y_t$) follow 
the renormalization group equations in Split Supersymmetry, which can be found in Ref.~\cite{Giudice:2011cg}~ at two loop level.\footnote{For earlier works see~\cite{Binger:2004nn}.}
 In numerical calculation, we evaluate $g_Y(M_S)$ and $g(M_S)$ by using the RGEs at 2-loop level and including the threshold corrections, $\Delta \lambda$, from sleptons and heavy Higgs boson (see Appendix A).

The threshold correction $\Delta \lambda$ includes the finite corrections induced by the trilinear coupling of stops, $A_t$. (However $A_t$ is small.) The corrections to $\lambda$ from sleptons and the Heavy Higgs boson are also included in $\Delta \lambda$. The detailed form of the threshold correction can be found in Ref.~\cite{Giudice:2011cg}. In our analysis, we evaluate $\lambda$ at the gaugino mass scale, by using the 2-loop RGEs in Split Supersymmetry and then the threshold correction from the Higgsino is included. (The Higgsino is heavier than the Wino.) The renormalization group evolutions of $\lambda$ as well as $Y_t$ below the gaugino mass scale are obtained by the RGEs of the SM. We evaluate the weak scale value of $\lambda(m_t)$ at two loop level.

\begin{figure}[t]
\begin{center}
\includegraphics[width=10cm]{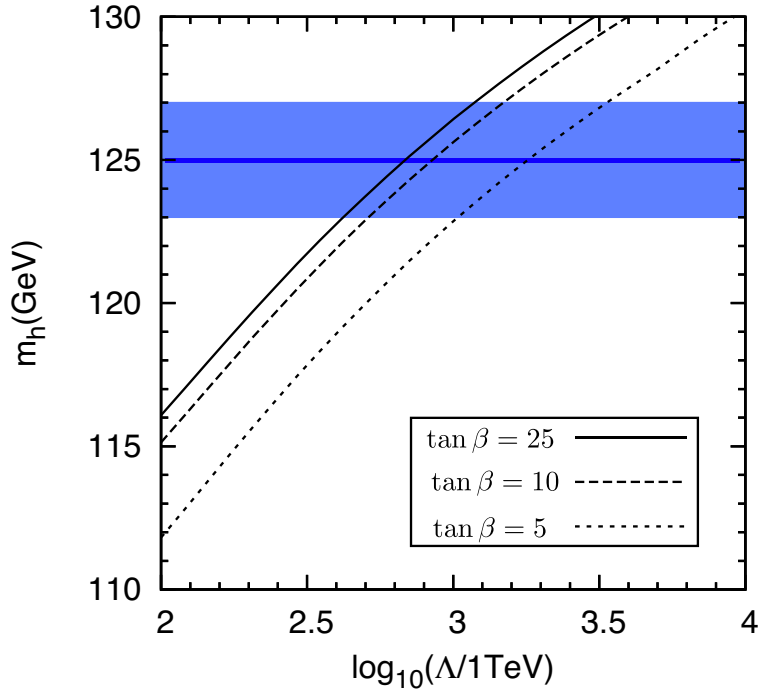}
\caption{The Higgs boson mass as a function of $\Lambda$ for various $\tan\beta$. We use the central values of the top mass and $SU(3)_C$ gauge coupling, $m_t=173.2\, {\rm GeV}$~\cite{Lancaster:2011wr} and $\alpha_S(m_Z)=0.1184$~\cite{PDG}. The messenger scale and the messenger number are taken as $M_{\rm mess}=2\Lambda$ and 
$N_{\rm mess}=4$. Here we assume that the gaugino masses are 1 TeV at the messenger scale.}
\label{fig:higgs_heavy}
\end{center}
\end{figure}

In Fig.~\ref{fig:higgs_heavy}, the Higgs boson masses for different $\tan\beta$ are shown as a function of $\Lambda$. We take the universal gaugino mass as $M_1=M_2=M_3=1$~TeV at the messenger scale, for simplicity. Since the quartic coupling given by Eq.~(\ref{eq:quartic}) is small for the small value of $\tan\beta$, e.g., $\tan\beta=5$, we need large radiative corrections to the quartic coupling; $\Lambda \sim 1500$~TeV is required for $m_h \simeq 125$ GeV. In the case that $\tan\beta$ is large enough, e.g., $\tan\beta \gtrsim 20$, the quartic coupling at the scale $M_S$ is large enough. 
Then, $\Lambda \simeq 400-1000$~TeV, which corresponds to $M_S \simeq 7-18$ TeV,
can explain the Higgs boson mass, $m_h \simeq 123-127$~GeV for $\tan\beta=25$. 

\subsection{Messenger contribution to the Higgs boson mass}
In low scale gauge mediation models, we have a new type of contribution to the Higgs boson mass which comes from
the messenger sector. We now discuss this contribution.

In the context of GGM, the low energy gauge coupling and the scalar (e.g., Higgs) 4-point coupling are given as
\beq
{\cal L}  &=& -\frac{1}{4 g^2_{(1)}}F_{\m\n}F^{\m\n}+\frac{g^2_{(0)}}{2}(\phi^\dagger t^a \phi)^2+\cdots 
\eeq
where $g_{(0)}$ and $g_{(1)}$ are given in terms of the UV gauge coupling $g$ as~\cite{Meade:2008wd}
\beq
g^2_{(1)}&\simeq&(1-g^2\tilde{C}_1(0))g^2, \\
g^2_{(0)}&\simeq&(1-g^2\tilde{C}_0(0))g^2.
\eeq
Therefore, contrary to the usual assumption in the MSSM, the gauge coupling and the Higgs 4-point coupling is not the same at the scale below the messenger scale.
The difference is given by
\beq
g^2_{(0)} &\simeq& g^2_{(1)}\left(1+g^2(\tilde{C}_1(0)-\tilde{C}_0(0)) \right) .
\eeq
Let us again use NDA. We may estimate $\tilde{C}_1(0)-\tilde{C}_0(0)$ as
\beq
\tilde{C}_1(0)-\tilde{C}_0(0) =\frac{AN}{8\pi^2},\label{eq:NDAforc1-c0}
\eeq
where $A$ is a numerical constant of order 1. This estimation is valid only in low-scale gauge mediation models ($F\sim M_{\rm mess}^2$),
because $\tilde{C}_1(0)-\tilde{C}_0(0)$ is typically suppressed by a factor $F^2/M_{\rm mess}^4$ if $F \ll M_{\rm mess}^2$.

Because the Higgs boson mass squared is proportional to $g_{(0)}^2$ (with $g^2$ taken as the sum of the $SU(2)_L$ and $U(1)_Y$ couplings), 
the correction to the Higgs boson mass is given as
\beq
\D m_h \simeq \frac{1}{16\pi^2} A N (g^2 \cos^2 \theta_W +g_Y^2 \sin^2 \theta_W)m_Z|\cos 2\beta|,
\eeq
where we have assumed the decoupling limit of the MSSM, i.e., the mass of the CP-odd neutral Higgs boson, $m_A$ is much larger than $m_h$.

In strongly coupled models, we may not be able to calculate the value of $A$.
However, the value of this parameter directly affects the Higgs boson mass.
The above correction is estimated as
\beq
\D m_h \sim 1~\GEV \cdot \left( A \cdot \frac{N}{5} \cdot |\cos 2\beta| \right). \label{eq:addcontribution}
\eeq 
If $A$ is somewhat larger than naively expected, this correction can raise the Higgs boson mass to $125~\GEV$ even if the scale $\Lambda$
is not so large.~\footnote{The Landau pole of the SM gauge coupling usually restricts the number of messengers $N$, but 
this constraint on the messenger number can be relaxed by high energy superconformal dynamics~\cite{Sato:2009yt}.
There is also a possibility that the effective messenger number increases drastically at the confinement scale, as in Ref.~\cite{Ibe:2010jb}.}
Note that this contribution just comes from the messenger sector, without any additional fields to the MSSM.

We should note, here, that there is an excellent possibility on the DM candidate in strongly coupled low scale gauge mediation models~\cite{Dimopoulos:1996gy,Hamaguchi:2007rb,Mardon:2009gw,Fan:2010is,Yanagida:2010zz}. In fact, some composite baryons in SUSY breaking sector or messenger
sector are stable or long-lived and hence they are good candidates for the DM.
%
%
In this scenario, the dynamical scale $\Lambda$ should be ${\cal O}(100)~\TEV$, and not ${\cal O}(1000)~\TEV$ for producing a correct DM density.
 (If the dark matter is strongly coupled in the early universe so that the unitarity bound of the (s-wave) annihilation cross section is almost saturated, 
the correct dark matter abundance may be obtained for the dark matter mass of order ${\cal O}(100)~\TEV$~\cite{Griest:1989wd}.)
If we do not have a contribution given by Eq.~(\ref{eq:addcontribution}), the scale $\Lambda$ is of order $1000~\TEV$ as discussed in
the previous subsection. Then the dark matter mass may also be of this order, although it is difficult to determine the precise value 
of the dark matter mass due to strong coupling. However, if $A$ is somewhat large, say $A \sim 3$, then
$\Lambda$ of order ${\cal O}(100)~\TEV$ may be sufficiently large to enhance the Higgs boson mass up to $125$ GeV.
Therefore, the scenario of strongly coupled hidden baryon dark matter may be possible if Eq.~(\ref{eq:addcontribution}) gives a significant 
contribution to the Higgs boson mass.

\subsubsection*{Example}
Here we give calculable examples of the parameter $A$.
First we consider weakly coupled messenger models.

Suppose that there is a vector-like pair of charged fields $\Psi$ and $\tilde{\Psi}$ in the representation $r$ 
(and its conjugate) of the $SU(5)_{\rm GUT}$,
respectively.
Let $\phi_1$ and $\phi_2$ be the mass eigenstates of the scalar fields in $\Psi=\phi+\cdots$ and ${\tilde \Psi}={\tilde \phi}+\cdots$,
that is,
\begin{eqnarray}
\left( 
\begin{array}{c}
\phi_1 \\
\phi_2
\end{array}
 \right)
=
\left( 
\begin{array}{cc}
\cos \theta & -\sin \theta \\
\sin \theta & \cos \theta 
\end{array}
 \right)
 \left( 
\begin{array}{c}
\phi \\
\tilde{\phi}^\dagger
\end{array}
 \right) ,
\end{eqnarray}
for a mixing angle $\theta$.
From the definition of $\tilde{C}_{s}~(s=0,1/2,1)$ given in Ref.~\cite{Meade:2008wd}, we find, after some calculation, that
\begin{eqnarray}
\frac{16\pi^2}{2t_r}\tilde{C}_0(0)&=&-\sin^2 2\theta
\left( \frac{m^2_{s1}}{m^2_{s1}-m^2_{s2}}\log \frac{m^2_{s1}}{\mu^2_R} - \frac{m^2_{s2}}{m^2_{s1}-m^2_{s2}}\log \frac{m^2_{s2}}{\mu^2_R} -1 \right) \nonumber \\
&&-\frac{1}{2}\cos^2 2\theta \left(\log \frac{m^2_{s1}}{\mu^2_R}+\log \frac{m^2_{s2}}{\mu^2_R} \right), \\
\frac{16\pi^2}{2t_r}\tilde{C}_1(0)&=&-\frac{2}{3}\log \frac{m^2_{f}}{\mu^2_R}-\frac{1}{6}\log \frac{m^2_{s1}}{\mu^2_R}--\frac{1}{6}\log \frac{m^2_{s2}}{\mu^2_R},
\end{eqnarray}
where $t_r$ is the Dynkin index of the representation $r$,
$m_{s1,2}$ are the masses of $\phi_{1,2}$, $m_f$ is the Dirac mass of the fermions in $\Psi$ and ${\tilde \Psi}$, and $\mu_R$
is a renormalization scale. We may identify $N_{\rm mess}=2t_r$, and the factor $A$ is defined as
\begin{eqnarray}
A=\frac{8\pi^2}{2t_r} (\tilde{C}_1(0)-\tilde{C}_0(0)).
\end{eqnarray}
One can check that the dependence on the renormalization scale $\mu_R$ is cancelled out in $A$.
One can also check that $A$ vanishes in the SUSY limit, $m_{s1}=m_{s2}=m_f$.

\begin{figure}[t]
\begin{tabular}{cc}
\begin{minipage}{0.5\hsize}
\begin{center}
\includegraphics[width=8cm]{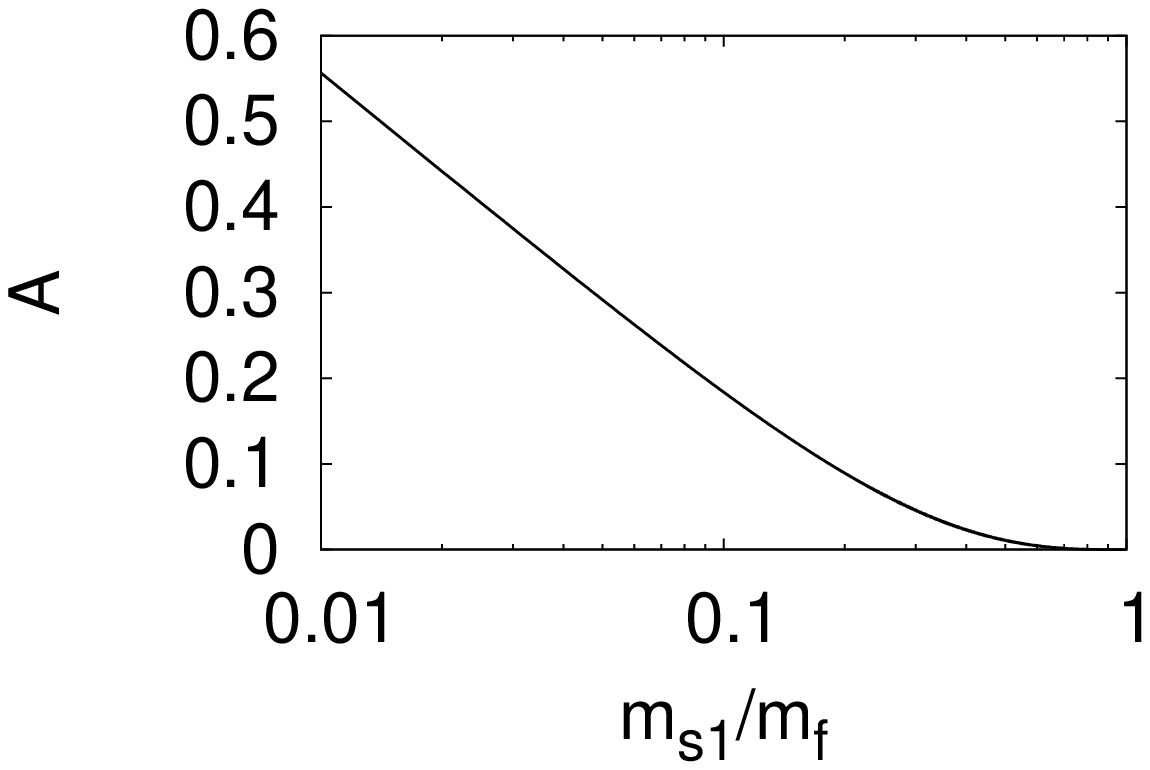}
(a) : $m_{s1}^2+m_{s2}^2=2m_{f}^2,~\theta=\pi/4$
\end{center}
\end{minipage}
\begin{minipage}{0.5\hsize}
\begin{center}
\includegraphics[width=8cm]{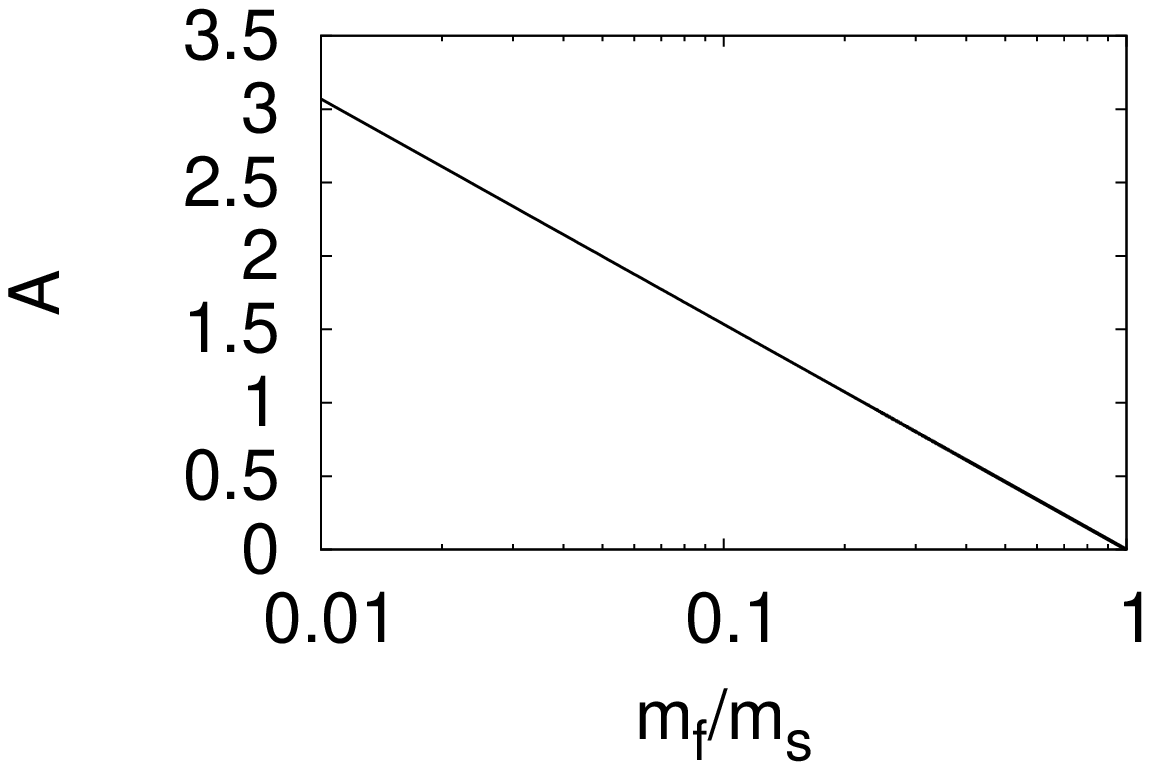}
(b) : $m_{s1}=m_{s2} \equiv m_s$
\end{center}
\end{minipage}
\end{tabular}
\caption{The factor $A$ in weakly coupled models. Left: the parameters are taken as $m_{s1}^2+m_{s2}^2=2m_{f}^2$ and $\theta=\pi/4$, which corresponds
to the minimal gauge mediation. The $A$ is plotted as a function of $m_{s1}/m_f$.
Right: the scalar masses are taken to be the same, $m_{s1}=m_{s2} \equiv m_s$. 
The $A$ is plotted as a function of $m_f/m_s$.}
\label{fig:Afactor}
\end{figure}

Two examples are shown in Fig.~\ref{fig:Afactor}.
In Fig.~\ref{fig:Afactor}-(a), we take $m_{s1}^2+m_{s2}^2=2m_{f}^2$ and $\theta=\pi/4$. This choice corresponds to the case of the minimal gauge mediation,
where $m^2_{s1}=M^2_{\rm mess}-yF,~m^2_{s2}=M^2_{\rm mess}+yF$ and $m^2_{f}=M^2_{\rm mess}$.
In Fig.~\ref{fig:Afactor}-(b), the scalar masses are taken to be the same, $m_{s1}=m_{s2} \equiv m_s$.

Next let us briefly discuss the case that a model is strongly coupled. Although it is generically very difficult to calculate $A$,
holographic duals may give some insight into strongly coupled gauge mediation~\cite{Benini:2009ff,McGuirk:2009am,Fischler:2011xd,McGuirk:2011yg,
Skenderis:2012bs,Argurio:2012cd}. For example, Refs.~\cite{Skenderis:2012bs,Argurio:2012cd} have obtained an ${\cal O}(1)$ (though negative) value
for $A$ in a specific holographic model.\footnote{To claim that $A$ is of order 1 in Refs.~\cite{Skenderis:2012bs,Argurio:2012cd},
we should define $N$ in Eq.~(\ref{eq:NDAforc1-c0}) as the contribution to the beta function in UV region, 
${\partial \o \partial \mu} (1/g^2)=-(N/8\pi^2)+({\rm MSSM~contribution})$. Then, for example,
the explicit expression for $\tilde{C}_{s}(0)~(s=0,1/2,1)$ in Ref.~\cite{Skenderis:2012bs} gives $A=-\frac{3}{8}(1+\ln2) \simeq -0.63$.}
It will be very interesting if we can find a holographic model in which $A$ is positive and somewhat large.

\section{Summary and Discussion}\label{sec:conc}
In this paper, we have considered two possible extensions of low scale gauge mediation models, which can accommodate the gravitino mass, $m_{3/2} \lesssim 16$ eV, and the 125 GeV Higgs boson mass: singlet extensions of the Higgs sector, and strongly coupled gauge mediation models. We have shown that the $Z_3$ invariant singlet extension, so-called NMSSM, is not suitable for explaining the 125 GeV Higgs boson, due to the Landau pole constraint. Then we have considered a more general model of the singlet extensions. In this case, the Higgs boson mass of 125 GeV can be explained without any difficulties. All dimension-full parameters in the superpotential are of the order of TeV. Although we have not discussed the origin of these dimension-full parameters,
they may be generated by using a $Z_3$ breaking spurion, $\left< \Phi \right>$, as $\left< \Phi \right>^2 S + \left<\Phi\right> S^2$, where $\left<\Phi \right> \sim 1$~TeV. 

In strongly coupled gauge mediation models, the SUSY breaking scale $\Lambda \simeq 400-1000$~TeV can explain the Higgs boson
mass $m_h \simeq 123-127$~GeV.
This $\Lambda$ is consistent with the very light gravitino, $m_{3/2} \lesssim 16$ eV. Although the squark mass is about 10 TeV,  gaugino masses may be as small as a few TeV. We have also discussed new effect to enhance the Higgs boson mass. This effect comes from the deviation of the D-term coefficient from 
that determined by the gauge couplings. 
Because of this effect, the Higgs boson mass may be enhanced by e.g., 3~GeV for $A\sim3,N \sim 5$ (see, Eq.(\ref{eq:addcontribution})). 
Although it may be challenging to obtain a large enhancement in this mechanism, 
the SUSY breaking scale $\Lambda$ can be as small as $\Lambda \sim 200$~TeV 
if the parameter $A$ is indeed large in strongly coupled models. 
In such a case, the composite baryon of the hidden strong gauge group may be a good candidate for the dark matter. 

\section*{Acknowledgment}
This work is supported by World Premier International Research Center Initiative
(WPI Initiative), MEXT, Japan,
by Grant-in-Aid for Scientific research
from the Ministry of Education, Science, Sports, and Culture (MEXT),
Japan, No. 22-7585(N.Y.) and No. 22244021 (T.T.Y.),
by JSPS Research Fellowships for Young Scientists (K.Y.).  

\appendix
\section{Threshold corrections to gauge couplings}
Here, we show the threshold corrections to the gauge coupling from sleptons and Heavy Higgs bosons:

\begin{eqnarray}
\Delta g_Y^2(M_S)&=& \frac{g_Y^4}{8\pi^2} \left( \frac{1}{2} \ln \frac{M_S}{m_{\tilde{L}}} + \ln \frac{M_S}{m_{\tilde{E}}} + \frac{1}{6}\ln \frac{M_S}{m_A}\right), \nonumber \\
\Delta g^2(M_S) &=& \frac{g^4}{8\pi^2} \left( \frac{4}{3} \ln \frac{M_S}{M_2} + \frac{1}{2} \ln \frac{M_S}{m_{\tilde{L}}} + \frac{1}{6} \ln \frac{M_S}{m_A}\right).
\end{eqnarray}



\end{document}